# Structural supermodulation enables resonant Raman determination of crystalline orientation in $Bi_2Sr_2CaCu_2O_{8+x}$ flakes

Flavia Lo Sardo,[1,2] Tommaso Confalone,[1,3] Shreya Kumbhakar,[1,3] Genda Gu,[4] Ahmad Omar,[1] Christian G.F. Blum,[1] Daniel Wolf,[1] Davide Massarotti,[5] Francesco Tafuri,[6] Axel Lubk,[1,7] Kornelius Nielsch,[1,3,2] Nicola Poccia,[1,6,*] and Golam Haider[1,†]

[1]*Leibniz Institute for Solid State and Materials Research Dresden (IFW Dresden), 01069 Dresden, Germany*
[2]*Institute of Materials Science, Technische Universitaet Dresden, 01062 Dresden, Germany*
[3]*Institute of Applied Physics, Technische Universitaet Dresden, 01062 Dresden, Germany*
[4]*Condensed Matter Physics and Materials Science Department, Brookhaven National Laboratory, Upton, NY 11973, USA*
[5]*Department of Electrical Engineering and Information Technology, University of Naples Federico II, 80125 Naples, Italy*
[6]*Department of Physics, University of Naples Federico II, 80126 Naples, Italy*
[7]*Institute of Solid State and Materials Physics, Technische Universitaet Dresden, 01062 Dresden, Germany*

$Bi_2Sr_2CaCu_2O_{8+x}$ (Bi-2212) is a *d*-wave superconductor with unconventional properties intrinsic to the material itself. These properties are of growing interest for quantum technologies in electronics and photonics, and twisted Josephson junctions can make direct use of them. However, Bi-2212 is strongly anisotropic, and it is challenging to find its crystal axes in flakes in advance, which limits device engineering. Here, we show that polarized resonant Raman spectroscopy reveals the orientation of both the *a* and *b* axes of Bi-2212 optically and non-destructively. By correlating Laue diffraction, HAADF-STEM and polarized Raman scattering on the same bulk crystal, we assign the *b* axis, along which the incommensurate structural supermodulation propagates, to the intensity maximum of the ∼27 $cm^{-1}$ supermodulation mode, which varies in counter-phase with the ∼118 $cm^{-1}$ Bi/Sr mode, along the *a* axis. Multi-wavelength measurements (458, 532, 660 nm) show that resulting in-plane contrast peaks at 532 nm and nearly vanishes at 660 nm, indicating that the anisotropy originates from an electronic resonance induced by the supermodulation. Using this resonance, we determine the crystallographic axes of exfoliated flakes and access their symmetry channels, providing a rapid orientation tool for cuprate van der Waals heterostructures.

## I. INTRODUCTION

High $T_c$ cuprate superconductors remain a central research topic in condensed matter physics, as their superconducting state emerges from an intricate interplay between lattice, charge and spin degrees of freedom [1–6]. Among them, the multi-element compound $Bi_2Sr_2CaCu_2O_{8+x}$ (Bi-2212) combines a $T_c$ of ∼91 K with a weakly bonded BiO–BiO double layer that allows the crystal to be cleaved down to a single unit cell, showing nodal *d*-wave superconductivity for pristine samples [7–11]. As a result, it has become a reference material for surface-sensitive spectroscopies and, more recently, a novel platform within the expanding field of twisted van der Waals quantum materials [12–16]. Its hole-doped complex crystal structure [17, 18], exhibits a small orthorhombic distortion, $(b - a)/a \approx 0.3\%$ [19], and carries an incommensurate structural supermodulation (SM) along the *b* axis, which arises from the lattice mismatch between the BiO and $CuO_2$ layers [20, 21]. In particular, the SM periodically displaces the atoms of the $CuO_2$ planes, modulating the local electronic density of states and the magnitude of the superconducting gap [22, 23]. Furthermore, it folds the electronic bands into superstructure replicas observed only along the modulation direction [24], and it breaks the in-plane equivalence of the *a* and *b* axes in both transport and optical response [20, 25, 26]. The hole-dopant oxygen, in turn, sets the carrier density and the spatial inhomogeneity of the superconducting state [17, 21, 27].

The electronic properties of Bi-2212 are therefore inseparable from its structural complexity, which at the same time, makes Bi-2212 fragile. The material loses oxygen [28] and easily degrades at the surface [29, 30], with detrimental effects on the $T_c$ and the Josephson coupling of devices [31, 32]. While almost everything known about the structure–property relationship of Bi-2212 has been established on bulk single crystals, mainly using X-ray diffraction [21, 33], scanning tunnelling microscopy [22], electron microscopy [21], angle-resolved photoemission [24] and Raman spectroscopy [34–43], direct structural investigations on thin flakes remain scarce. Yet, in mechanically exfoliated flakes, phenomena such as the alteration of the SM coherence [21] or oxygen loss during fabrication can fundamentally modify the material's properties [8, 30, 44]. Therefore, whether an exfoliated flake retains the anisotropy of the bulk crystal, and along which direction, cannot simply be assumed.

Knowing this aspect is important because exfoliated flakes are the building blocks of the emerging field of cuprate twistronics [6, 13], spanning from their integration into twisted Bi-2212 Josephson junctions [4, 12,

* nicola.poccia@unina.it
† g.haider@ifw-dresden.de

14, 45–47] to their incorporation into printed circuit boards and microwave circuits for quantum hardware [31, 48, 49]. In many of these devices, the relevant physics is governed by the orientation of the *d*-wave order parameter relative to the junction interface and to the partner flake. Because the orthorhombic *a* and *b* axes run along the nodal directions of the *d*-wave gap, the in-plane crystallographic axes directly correlates to the nodal geometry of a device. The absolute assignment of *a* and *b*, and not only the relative twist angle, therefore determines the in-plane anisotropy of both transport [25] and optical response [26]. Identifying these axes in an exfoliated flake is, however, surprisingly difficult. Laue and X-ray diffraction, the standard tools for bulk crystals, lack the scattering volume provided by a thin flake on a substrate [50, 51]. Diffuse scattering from the SM, disorder and strain renders the distinction between *a* and *b* ambiguous, even in bulk crystals [21]. Cross-sectional scanning transmission electron microscopy (STEM) resolves the SM directly [21, 32] but is destructive, and scanning tunnelling microscopy requires atomically clean surfaces cleaved in situ [22]. Flake edges, finally, do not reliably follow a given crystallographic direction. Consequently, only a few studies have addressed exfoliated Bi-2212 flakes with optical probes, and they have focused on phonon assignment, optical contrast and ambient stability [30, 40]. We recently showed that the polarization dependence of the Bi/Sr Raman mode can be used to quantify the relative twist angle in Bi-2212 homojunctions [52]. That approach, however, measures relative angles only and does not distinguish the axes. Moreover, the origin of such a pronounced Raman anisotropy has remained unexplained. The low-frequency Raman modes activated by the SM are the natural spectroscopic fingerprint of the *b* axis [34, 53], yet a polarization-resolved study of these excitations has not been reported.

Here, we close this gap by correlating Laue diffraction, polarized Raman scattering and high-angle annular dark-field (HAADF) STEM on the same crystal to distinguish *a* and *b* axes. Together, these measurements establish that the *b* axis coincides with the intensity maximum of the SM mode and with the intensity minimum of the Bi/Sr mode, which oscillates in counter-phase. With this knowledge we determine the the axes of exfoliated flakes and use this orientation to access the distinct symmetry channels of the $CuO_2$ planes. Excitation-energy-dependent measurements at 458, 532 and 660 nm reveal the origin of the polarization contrast. It is strongest at 532 nm, redistributes among different phonons at 458 nm, and nearly vanishes at 660 nm. The SM therefore imprints its direction on the resonant electronic intermediate states, and through them on the Raman tensor. The resulting method is rapid and non-destructive, compatible with flakes already placed on device substrates, and requires only a standard confocal Raman microscope.

## II. RESULTS AND DISCUSSION

A schematic representation of the Bi-2212 unit cell is illustrated in Fig. 1a, where the interatomic bonds are represented by the black lines. The in-plane crystallographic axes are denoted as *a*, *b* and *c*, with lattice parameters of $a = 5.401$ Å, $b = 5.397$ and $c = 30.78$ Å [54]. Along the *b*-axis, the material exhibits an incommensurate structural SM driven by the lattice mismatch between the $CuO_2$ and BiO atomic planes [20]. Laue diffraction patterns were acquired on a single-crystal bulk Bi-2212 sample (thickness $\sim$100 $\mu$m), with the *c*-axis out of plane, directly identified due to natural cleaving/exfoliation (Fig. 1b). The resulting pattern is highly diffused, mainly due to a combination of incommensurate structural SM, mesoscopic disorder and strain, making a direct indexing of the crystallographic axes infeasible [21]. Given that the *a*, *b* lattice parameters are rather similar, a four-fold symmetry could be outlined, although quite diffused. Additional symmetric directions are noted at 45° as well and are highlighted in different colors. One of these orientation directions was arbitrarily chosen as the 0° reference and is highlighted in red. The remaining directions are consequently assigned to 45°, 90°, and 135° and are indicated by purple, brown, and blue lines, respectively.

The 0° direction was used as a reference for the subsequent Raman measurements and was marked directly on the sample holder. This provided a macroscopic angular reference that allowed the crystal to be reproducibly oriented with respect to the polarization direction of the Raman setup, as represented in the schematization of Fig. 1c.

It is already established that polarization-dependent Raman measurements on Bi-2212 can reveal a strong angular dependence of the intensity for specific phonon modes [36, 52]. Here, we exploit this mode anisotropy to establish a connection with the directions identified in the Laue diffraction pattern. For this purpose, a polarized Raman measurement from 0° to 180° was performed, with the 0° polarization angle defined along the 0° reference direction derived from the Laue pattern.

Two Raman modes were selected for the analysis, namely the modes at $\sim$27 $cm^{-1}$ and $\sim$118 $cm^{-1}$, as shown in Fig. 1d. The acoustic phonon mode at $\sim$27 $cm^{-1}$ is associated with the fundamental SM of Bi-2212 [34, 53], which breaks the equivalence between the two in-plane crystallographic directions *a* and *b* [20]. The mode at $\sim$118 $cm^{-1}$ is associated with vibrations involving the heavy Bi [37–40] or Sr atoms [35, 36, 40, 41, 43] and exhibits a significantly stronger Raman signal than the mode at $\sim$27 $cm^{-1}$.

Although previous studies have reported that polarized Raman spectra measured along the *a* and *b* crystallographic directions exhibit differences in both peak presence and intensity [34], a quantitative relationship between these features has not always been explicitly established. More specifically, when the intensity of the $\sim$118

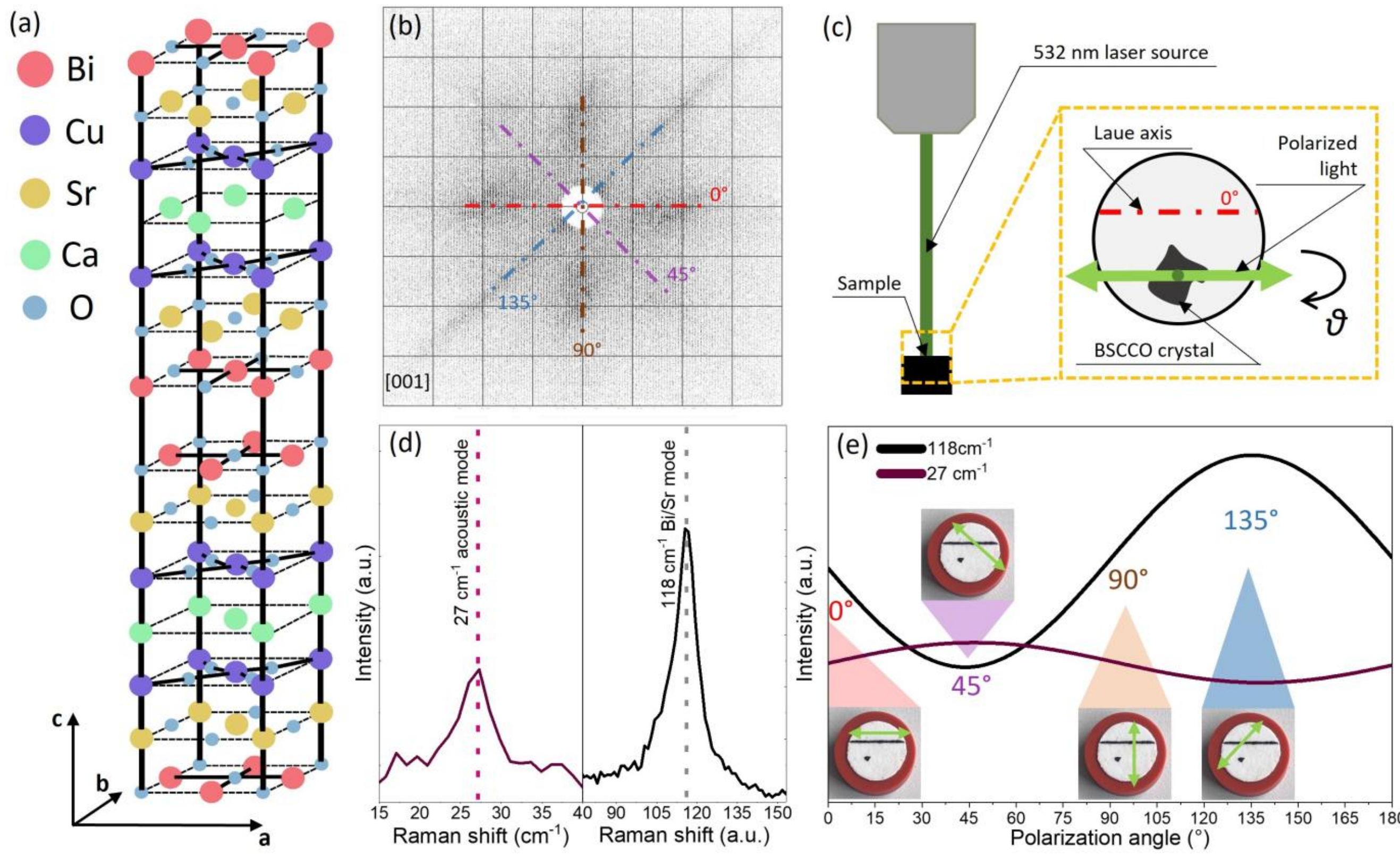

FIG. 1. (a) Schematic of Bi-2212 unit cell; (b) Laue diffraction pattern of a bulk Bi-2212 single crystal. The colored dotted lines define the angular reference frames at 0°, 45°, 90°, and 135° (indicated in red, purple, brown, and blue, respectively); (c) Simplified schematic of the polarized Raman experimental geometry. The dash-dotted red line marks the 0° reference axis determined by Laue diffraction. The double-headed green arrow represents the in-plane polarization direction of the incident laser beam, while $\vartheta$ denotes the rotation of the polarized light; (d) Polarized Raman spectra highlighting the $\sim$27 cm$^{-1}$ acoustic mode and $\sim$118 cm$^{-1}$ *c*-axis vibration of Bi/Sr; (e) Angular dependence and corresponding sinusoidal fits (see eq. 2, Materials and Methods Section) of the $\sim$27 cm$^{-1}$ and $\sim$118 cm$^{-1}$ Raman mode intensities as a function of the polarized light angle $\vartheta$. The two modes exhibit a clear counter-phase modulation. The photographic insets illustrate the actual orientation of the sample utilized for Laue measurements at the Raman microscope stage with respect to the fixed laser polarization (green arrow) at representative angles.

cm$^{-1}$ mode is lower, additional low-frequency peaks at $\sim$27 cm$^{-1}$ and 52 cm$^{-1}$ become visible [34, 38]. These modes are considered characteristic signatures of the SM, with the one at $\sim$52 cm$^{-1}$ being a second-order amplitude mode of the structural SM. Since the incommensurate SM occurs along the *b*-axis [20, 21, 34], the direction along which these additional features are observed can be assigned to the *b* crystallographic axis. Conversely, when the intensity of the $\sim$118 cm$^{-1}$ mode is higher, these low-frequency peaks are no longer visible, corresponding to the *a*-axis direction. The maximum intensity of the $\sim$118 cm$^{-1}$ mode along the *a*-axis indicates that its Raman tensor is dominated by the $R_{aa}$ component ($R_{aa} > R_{bb}$).

Here, we use the anisotropy of the modes at $\sim$118 cm$^{-1}$ and $\sim$27 cm$^{-1}$ to track the angular evolution of the Raman signal over 180° and to further quantify this relationship.

The fits (see Eq. 1-3) to the angular dependence of the two modes are shown in Fig. 1e (See S1, Supporting Information (SI) for the experimental data). It is also shown in the photographic insets how the sample was oriented with respect to the polarized light. At 0°, the intensities of both the $\sim$27 cm$^{-1}$ and $\sim$118 cm$^{-1}$ modes lie approximately halfway between their respective maximum and minimum values. At 45°, however, the $\sim$27 cm$^{-1}$ mode reaches its maximum intensity (45° $\pm$ 4.7°), while the $\sim$118 cm$^{-1}$ mode reaches its minimum (45 $\pm$ 1.1°). Based on the relationship described above, this observation suggests that the polarization direction at 45° corresponds to the *b*-axis of the crystal. At 90°, both modes again exhibit intermediate intensity values. Finally, at 135°, the $\sim$27 cm$^{-1}$ mode reaches its minimum, while the $\sim$118 cm$^{-1}$ mode reaches its maximum intensity, corresponding to the *a*-axis direction. Overall, the two modes exhibit an opposite angular dependence, with the maximum of one mode corresponding to the minimum of the other thereby clarifying the counter-phase symmetry relationship that couples them.

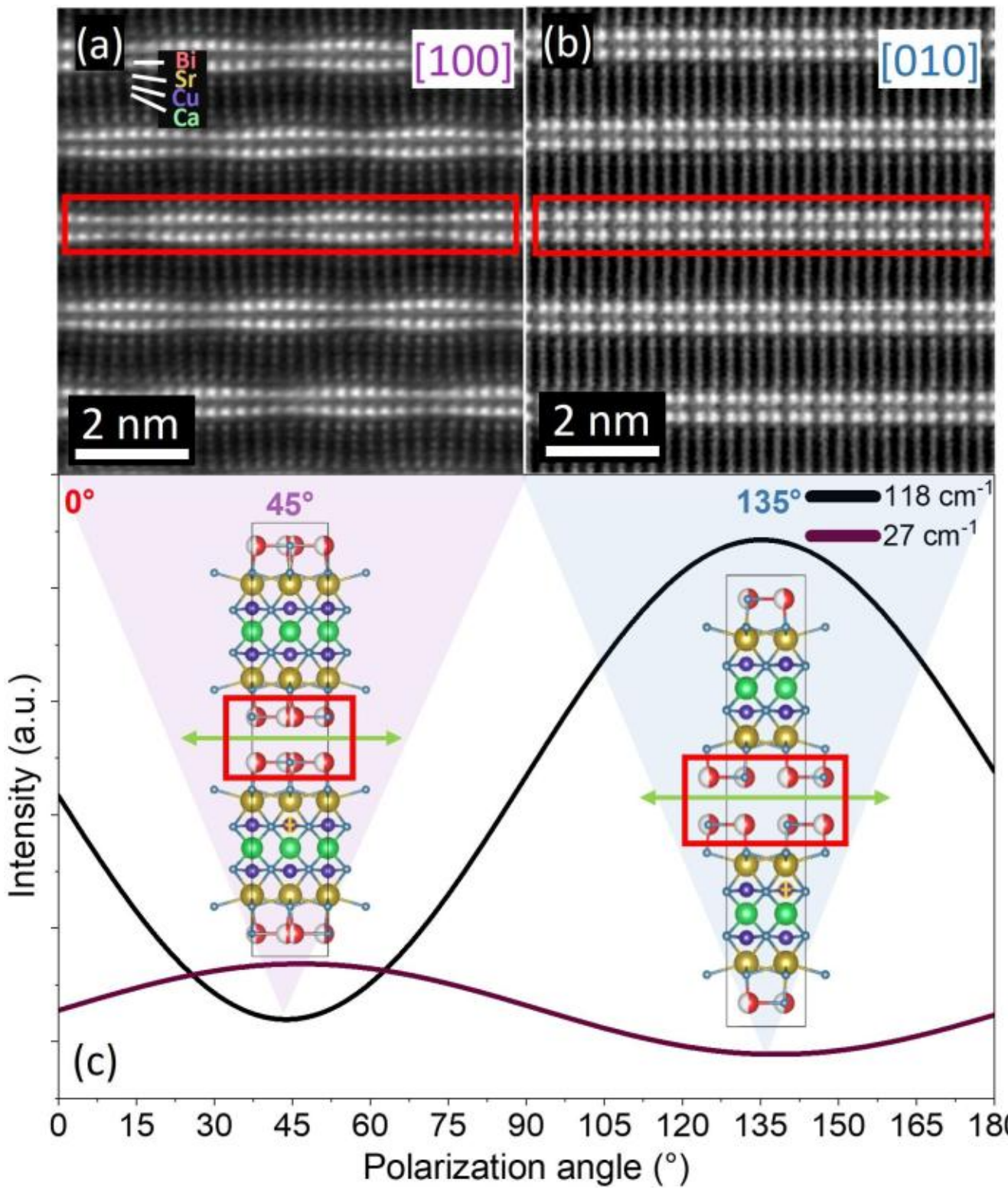

FIG. 2. (a) HAADF-STEM image of the Bi-2212 crystal structure with view along the [100] zone axis obtained from a 45° cross-sectional cut. The red rectangle highlights the periodic, wave-like SM of the BiO atomic layers. The brightest spots are identified with Bi atoms; (b) HAADF-STEM image of the same crystal with view along the [010] zone axis (cross sectional cut at 135°). The red rectangle highlights the Bi atoms along the *a*-axis; (c) Angular dependence and corresponding sinusoidal fits of the $\sim$27 cm$^{-1}$ acoustic mode (purple) intensity due to the SM along *b* and $\sim$118 cm$^{-1}$ (black) *c*-axis vibration due to the vertical modulation.

In order to validate our approach, the same sample was subsequently investigated by HAADF-STEM image. Two cross-sectional lamellae were prepared: one cut at 45° and the second at 135° with respect to the Laue reference direction. The corresponding HAADF-STEM images are shown in Fig. 2a and b. The two images clearly exhibit different structural features. In particular, the image acquired at 45° (Fig. 2a, along the [100] zone axis) reveals the presence of the SM [32, 45], recognizable from the characteristic wave-like pattern of the BiO layers, highlighted in the red rectangle. On the other hand, at 135° (Fig. 2b, along the [010] zone axis) no such modulation is observed. Instead, the red rectangle highlights BiO layers that appear linear, displaying a regular, unmodulated atomic spacing.

These observations are directly correlated with the angular dependence obtained from the Raman measurements. As shown schematically in Fig. 2c, the direction at 45°, corresponding to the maximum intensity of the $\sim$27 cm$^{-1}$ mode and the minimum intensity of the $\sim$118 cm$^{-1}$ mode, coincides with the direction along which the SM is observed by HAADF-STEM. This independently confirms the assignment of the 45° direction to the *b*-axis. Conversely, the 135° direction does not exhibit the characteristic SM but reveals an alternating short-long spacing of the Bi atoms which directly modulates the lattice polarizability along the transversal direction, corresponding to the *a*-axis. This electronic transition is maximal along the ordered *a*-direction because of the absence of SM-induced disorder, and reduces along the *b*-direction where the modulation perturbs the lattice. This behavior is exactly inverse to the acoustic phonon excitation of the supermodulation at $\sim$27 cm$^{-1}$, which is minimal along the ordered direction *a* and maximal along the modulated direction *b*, directly reflecting in the observed counter-phase intensity maximum of the $\sim$118 cm$^{-1}$ mode. Further crystallographic details of these projections, their simulated electron diffraction pattern and their corresponding Fourier Transforms (FTs) are provided in the SI (Figs. S2 and S3).

Finally, we want to extend this approach to exfoliated Bi-2212 flakes. A flake with a thickness of $\sim$100 nm was selected, and an angular-resolved polarization measurement was performed (Fig. S4). By isolating the $\sim$118 cm$^{-1}$ mode [52] and tracking its periodic angular dependence, the minimum intensity position (corresponding to the *b*-axis direction) was identified at 14° ($\pm$ 1.1°), while the maximum intensity position (corresponding to the *a*-axis direction) was located at 104° ($\pm$ 1.1°). These orientation parameters were subsequently used to align the system and perform Raman measurements in three distinct symmetry channels. Figure 3a illustrates the exfoliated flake used in this study along with the parallel polarizer-analyzer configurations, aligned along the *y*-axis (14°) and *x*-axis (104°) (*x* and *y* are defined to be along the *a* and *b* crystallographic axes), corresponding to the $A_g$ irreducible representation. The resulting Raman spectra are displayed in Fig. 3b.

Conversely, the crossed symmetry channels were probed along $y$ = 14° and $x$ = 104° (Fig. 3c), whereas the other crossed channel was accessed by rotating the optical axes by 45°, corresponding to projection angles of 59° ($y'$) and 149° ($x'$) (Fig. 3e). Physically, the $xy$ symmetry channel selectively probes the anti-nodal directions of the $CuO_2$ plane since for our convention the $x$ and $y$ lie along the crystallographic $a$ and $b$ axes, which are rotated by 45° with respect to the Cu–O bonds. This configuration corresponds to the $B_{1g}$ irreducible representation.The $x'y'$ symmetry channel targets the nodal directions, corresponding to the $B_{2g}$. In the $xy$ configuration (Fig. 3d), the mode at approximately 285–290 cm$^{-1}$ is clearly resolved. This specific feature is widely associated with the out-of-phase buckling vibration of oxygen atoms within the $CuO_2$ planes [55, 56]. Conversely, the $x'y'$ configuration (Fig. 3f) exhibits significantly weaker

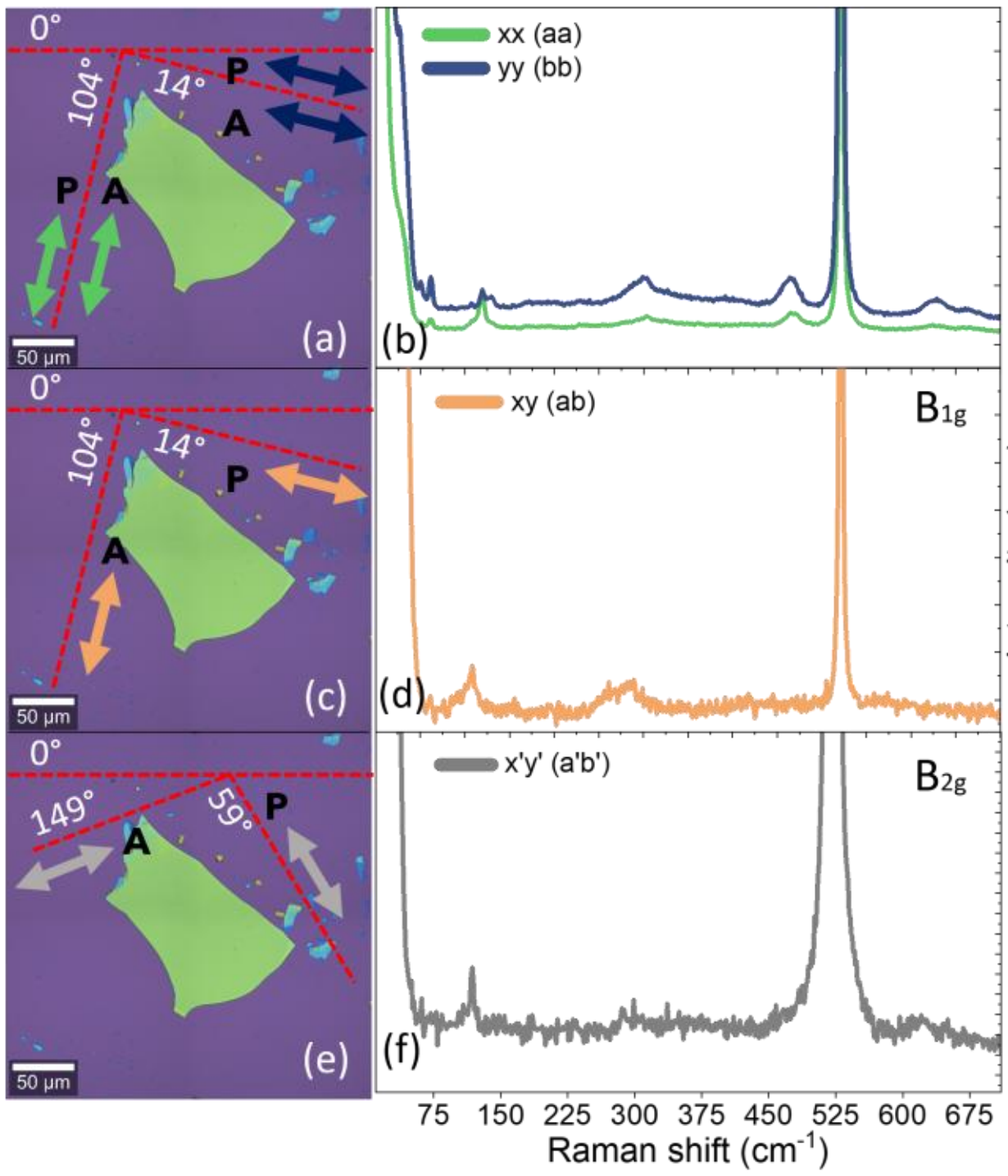

FIG. 3. (a, c, e) Optical microscope images of the exfoliated Bi-2212 flake (thickness ∼100 nm) showing the polarizer (P) and analyzer (A) configurations selected to isolate the different symmetry channels relative to the crystal axes. (a) Parallel geometries aligned at 14◦ (*yy*) and 104◦ (*xx*); (c) Crossed geometry aligned at 14◦/104◦ (*xy*); (e) Crossed geometry rotated to 59◦/149◦ ($x'y'$); (b, d, f) Polarized Raman spectra acquired at a 532 nm excitation wavelength within the isolated symmetry channels: (b) Raman spectra related to parallel configurations *xx* and *yy*, (d) crossed configuration *xy*, and (f) crossed configuration $x'y'$.

Raman features, confirming the strong anisotropy of the electronic and structural responses in the exfoliated sample.

Having established the correspondence between the polarization dependence of the Raman modes and the in-plane crystallographic axes, we now address whether this anisotropy reflects only the static lattice symmetry or is additionally controlled by the electronic states involved in the scattering process. A closer look at the anisotropy of the modes at ∼27 cm$^{-1}$, ∼52 cm$^{-1}$, ∼61 cm$^{-1}$, ∼118 cm$^{-1}$ is shown in the color plots of Fig. 4a-c. We note the mode at ∼27 cm$^{-1}$ (Fig. 4a) exhibits the same angular dependence as the ∼52 cm$^{-1}$ mode (Fig. 4b), both going in counter-phase with respect to the ∼118 cm$^{-1}$ mode (Fig. 4c). We further observe that the ∼52 cm$^{-1}$ mode appears as a shoulder of a stronger peak at approximately ∼61 cm$^{-1}$ which does not exhibit any significant angular dependence (Fig. 4b).

To discern the underlying mechanism, the excitation-energy-dependent measurements shown in Fig. 4d–f and Table I provide a clear distinction regarding the relative contributions of the static lattice symmetry versus the electronic modulation.

As noted in Fig. 4d and Tab. 1, upon excitation at 458 nm, the ∼52 and ∼118 cm$^{-1}$ modes exhibit a moderate anisotropy (19% and 40%), whereas the ∼61 and ∼460 cm$^{-1}$ modes develop a clear polarization dependence. At 532 nm, the polarization contrast becomes highly pronounced for the ∼27, ∼52 and ∼118 cm$^{-1}$ modes, with anisotropies of 95%, 54% and 65%, respectively (Fig. 4e, Tab. 1). In contrast, the ∼61 and ∼460 cm$^{-1}$ modes remain nearly isotropic (5% and 3%). Finally, at 660 nm, the directional response is almost entirely suppressed, and only a weak residual anisotropy (9%) of the ∼118 cm$^{-1}$ mode remains (Fig. 4f, Tab. 1). If the mode anisotropy depended only on the vibrational modes of the crystal structure, varying the excitation energy would modify the overall scattering intensity alone while preserving the same geometric and directional trends across all wavelengths. Instead, altering the laser photon energy fundamentally rearranges the anisotropic behavior of specific modes, causing their polarization dependence to either shift or completely vanish.

In particular, at 532 nm wavelength, the ∼61 and ∼460 cm$^{-1}$ modes show anisotropies of only a few percent which is consistent with the minor orthorhombic distortion of Bi-2212 ($(b - a)/a \approx 0.3\%$) [19]. On the other hand, the intensities of the ∼118 cm$^{-1}$ and ∼27 cm$^{-1}$ modes differ between the two in-plane axes by factors of ∼5 and ∼40, respectively. The anisotropy of the 27 cm$^{-1}$ mode was discussed previously and is directly explained by the propagation of the SM along the *b*-axis. For the ∼118 cm$^{-1}$ mode, a static lattice anisotropy of just ≈0.3% cannot account for such a pronounced contrast. Given that its displacement pattern is predominantly out of the *ab*-plane, its in-plane tensor components, $R_{aa}$ and $R_{bb}$, are determined not by the direction of atomic motion but by how the vibration modulates the electronic polarizability. The anisotropy must therefore reside in the electronic part of the Raman vertex and be directly enhanced by resonance effects.

In the resonant regime, occurring when the laser photon energy matches a real electronic transition of the material, the Raman tensor becomes intrinsically dependent on the excitation energy, because virtual electronic intermediate states mediate the phonon scattering [42, 57]. In Bi-2212, it was found that the resonant response is highly sensitive to both the electronic character and spatial location of the phonon-induced perturbation, as evidenced by the distinctly different resonance profiles of $CuO_2$ plane oxygen vibrations (attributed to electronic transitions between hybridized Cu 3d – O 2p states and the upper Hubbard band) compared to those of BiO and bridging-oxygen layers [24, 42], as schematically illustrated in Fig. 4g. Changing the incident photon energy modifies both the optical matrix elements connecting the ground state to these intermediate states and the electron–phonon matrix elements through which a given vibration perturbs them. The ratio $R_{aa}/R_{bb}$ is therefore not fixed by symmetry alone but also depends on which intermediate states are brought into resonance. The tensor ratio $|R_{aa}/R_{bb}|$ of the $\sim$118 cm$^{-1}$ mode is $\approx$1.5, 2.2 and 1.1 at 458 nm (2.71 eV), 532 nm (2.33 eV) and 660 nm (1.88 eV), respectively. This non-monotonic evolution, peaking at intermediate photon energy, is characteristic of a resonance rather than a structural selection rule.

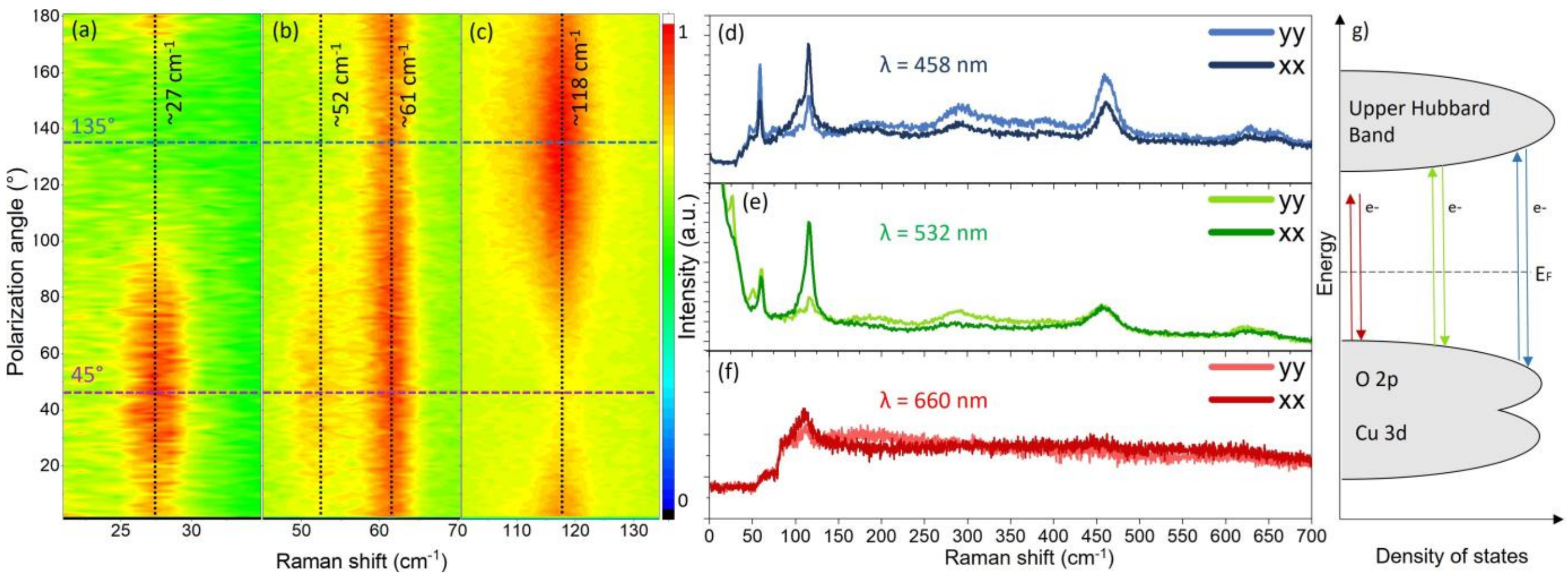

FIG. 4. (a–c) Color maps displaying the intensity profiles of the isolated low-frequency Raman modes as a function of the polarization rotation angle over a 180° range, recorded with a 532 nm excitation wavelength; (a) The $\sim$27 cm$^{-1}$ SM-driven mode, showing maximum intensity at 45° (purple dashed line) and near-complete suppression at 135° (blue dashed line); (b) Coexisting spectral windows of the $\sim$52 cm$^{-1}$ and $\sim$61 cm$^{-1}$ excitations; the former mimics the angular response of the $\sim$27 cm$^{-1}$ peak, whereas the structural $\sim$61 cm$^{-1}$ mode exhibits nearly isotropic behavior; (c) The $\sim$118 cm$^{-1}$ vibrational mode, displaying a pronounced angular modulation in counter phase to the $\sim$27 and $\sim$52 cm$^{-1}$ excitations; (d–f) Polarized Raman spectra under parallel configuration geometries *xx* and *yy* channels) at three distinct laser excitation wavelengths ($\lambda$): (d) 458 nm (blue), (e) 532 nm (green), and (f) 660 nm (red). Higher laser excitation energies selectively enhance the polarization anisotropy of the higher-frequency vibrational modes, whereas lower excitation energies shift the resonance enhancement toward the lower-frequency regime; (g) Schematic electronic energy diagram and density of states of Bi-2212 illustrating the multi-wavelength resonant scattering mechanism. The blue, green, and red arrows depict interband optical transitions between the hybridized Cu 3d–O 2p valence states and the upper Hubbard band for different $\lambda$.

This effect is particularly significant in Bi-2212 because the incommensurate SM explicitly breaks the equivalence of the two in-plane directions [20]. The periodic lattice distortion along *b* reconstructs the local electronic structure and generates superstructure-derived bands through folding and hybridization, which are observed by ARPES only along the modulation direction [24]. Optical transitions involving these states therefore acquire a strongly directional character: they are accessible for light polarized along *b*, but not along *a*. When the excitation energy couples efficiently to such SM-sensitive intermediate states, the resonant Raman susceptibility inherits their in-plane asymmetry. The result is a large imbalance between the effective *a* and *b*-axis tensor components. The response at 532 nm illustrates this mechanism most clearly. The $\sim$27 and $\sim$52 cm$^{-1}$ modes are Raman active only through the SM itself, as the fundamental and second-order excitations of the modulation. Their scattering amplitude is thus intrinsically tied to the *b*-polarized, SM-reconstructed pathway. This is consistent with their maximum intensity along *b*, as independently established by HAADF-STEM, and with the near-complete suppression of the $\sim$27 cm$^{-1}$ mode along *a*. The $\sim$118 cm$^{-1}$ mode, located in the Bi/Sr layers where the SM displacements are largest, couples to the same anisotropic intermediate states but with the opposite weighting of the two in-plane components. It therefore peaks along *a*, in counter-phase with the SM modes. Meanwhile, the $\sim$61 cm$^{-1}$ mode remains almost isotropic under the same conditions, which demonstrates that the 532 nm resonance acts selectively on specific phonon channels rather than uniformly on the spectrum.

The behavior at 458 nm provides a complementary result. Here the anisotropy is distributed across a broader set of phonons. The $\sim$61 and $\sim$460 cm$^{-1}$ modes, nearly isotropic at 532 nm, acquire appreciable directional contrast, whereas the anisotropy of the SM-sensitive $\sim$52 and $\sim$118 cm$^{-1}$ modes is reduced. This suggests that the higher excitation energy accesses a different manifold of intermediate states, likely with larger weight in the SrO and apical-oxygen layers, whose couplings involve a wider range of atomic vibrations. In line with earlier resonance-profile studies [42], a mode's resonant response reflects the spatial overlap of its displacement pattern with the resonant electronic states, not its vibrational frequency. The excitation energy thus controls which phonon channels project the underlying electronic anisotropy into the Raman response.

| Assignment | Mode ($cm^{-1}$) | % of anisotropy ($r$) | | |
|---|---|---|---|---|
| | | 458 nm | 532 nm | 660 nm |
| SM | ∼27 | - | 95 | - |
| $2^{nd}$ order SM | ∼52 | 19 | 54 | - |
| Bi | ∼61 | 25 | 5 | - |
| Sr/Bi | ∼118 | 40 | 65 | 9 |
| $O_{Bi}$ | ∼290 | 25 | 31 | - |
| $O_{Sr}$ | ∼460 | 22 | 3 | - |

TABLE I. Summary of the experimentally obtained % of anisotropy ($r$) for the extracted Bi-2212 Raman modes under blue (458 nm), green (532 nm), and red (660 nm) laser excitations.

Finally, the near disappearance of the polarization contrast at 660 nm provides an important reference condition. The ∼118 $cm^{-1}$ phonon remains clearly Raman active while its anisotropy collapses, showing that the vibration itself persists but the strongly directional enhancement is lost. This separation between phonon visibility and polarization contrast strongly suggests that the large anisotropy at 532 nm originates from resonant electronic enhancement rather than from a purely structural selection rule. As the excitation energy moves away from the resonant intermediate states, the tensor relaxes toward the nearly tetragonal limit set by the small orthorhombic distortion.

Taken together, these results identify the SM as more than a source of additional phonon modes. It acts as a periodic anisotropic potential that reshapes the electronic states mediating Raman scattering, so that the resonant Raman tensor carries a spectroscopic fingerprint of the modulation direction. Among the excitation energies studied, 532 nm maximizes this electronic–structural coupling for both the ∼27 and ∼118 $cm^{-1}$ modes. It therefore provides the highest angular contrast, and hence the most reliable optical determination of the crystallographic axes, including in thin flakes where only the ∼118 $cm^{-1}$ mode is readily accessible.

## III. CONCLUSION

The incommensurate structural SM along the *b*-axis emerges as a central factor driving the polarization anisotropy observed in the low-frequency Raman modes of Bi-2212. By establishing a correlation framework that benchmarks polarized Raman scattering against structural data from Laue diffraction and HAADF-STEM on bulk single crystals, we mapped the relationship between optical selection rules and real-space crystallographic orientations. In addition, because the lattice SM acts as a periodic modulation potential, it introduces a strong directional anisotropy into the electronic intermediate states and interband transitions of the $CuO_2$ planes. By using green laser light (532 nm), the anisotropy of the low frequency modes related to SM amplifies. Under blue light (458 nm), the anisotropy is still observed but is weaker compared to the green excitation, while under red light (660 nm) it is almost completely lost. Leveraging this resonant Raman enhancement, we successfully transferred these selection rules to mechanically exfoliated flakes, establishing a robust, non-destructive, and purely optical methodology to rapidly determine the in-plane crystallographic axes of Bi-2212 flakes using exclusively polarized Raman measurements.

## IV. MATERIALS AND METHODS

*Sample preparation:* A BSCCO bulk single crystal ($T_c$ ∼91 K, $\Delta T_c$ ∼2 K [44, 49]) was used for Laue diffraction, Raman spectroscopy, and TEM measurements, while an exfoliated flake on a $Si/SiO_2$ substrate was used for symmetry resolved Raman spectroscopy measurements.

For the HAADF-STEM, cross-sectional lamellae were extracted from the bulk crystal and thinned down using a focused ion beam (FIB) system before being transferred onto a standard HAADF-STEM grid.

*Spectral acquisition:* Raman spectra at 532 nm wavelength were acquired in a polarized backscattering geometry at room temperature using a WITec Alpha 300 confocal Raman microscope.

Raman spectra at 458 nm and 660 nm wavelength were acquired in a backscattering geometry at room temperature using a LabRAM HR Evolution, HORIBA Scientific confocal Raman microscope.

Both setups were equipped with a 10x, 50× and 100x objective lenses of respectively NA = 0.25, NA = 0.75 and NA = 0.9 and a diffraction grating with 1800 lines/mm, yielding a spectral resolution of approximately 1.2 $cm^{-1}$.

For the polarization dependent measurements on bulk like crystal, the laser power was adapted according to the laser wavelength to not burn the sample. An acquisition time of 3 s per spectrum, repeated over 181 points (i.e., from 0° to 180°) was used.

A ∼3 mW laser power with 60 s integration time and 3 accumulations was used for The single spectra on the exfoliated flake.

*Laue:* Laue diffraction was performed using a real-time back reflection Laue camera system (Multiwire MWL 120 with Northstar software) using the white light from a tungsten X-ray tube. For the measurements, the bulk crystal was fixed onto a custom-designed adhesive support to ensure stable angular positioning.

*HAADF-STEM:* High-angle annular dark-field scanning transmission electron microscopy (HAADF-STEM) was performed on cross-sectional TEM lamellae prepared by focused-ion-beam (FIB) milling. An aberration-corrected FEI Titan 80-300 TEM (Thermo

Fisher, USA) operating at 300 kV was used. For HAADF-STEM imaging, a semi-convergence angle of 21 mrad and a camera length of 91 mm (yielding a collection-angle range of 64-200 mrad) were selected. A Wiener filter was applied to denoise the HAADF STEM images.

*Raman data analysis:* Data were analyzed using Origin software for both the representation of Raman spectra and the evaluation of periodic trends and angular shifts. The raw data of the anisotropic modes were fitted with a sinusoidal function of the form:

$$y = y_0 + A\sin\left[\pi\frac{(x - x_c)}{w}\right] \quad (1)$$

where $y_0$ is the baseline, A the amplitude, $x_c$ the phase shift, and w half the oscillation period. The maxima were calculated from the fit parameters as:

$$x_{max} = x_c + \frac{w}{2} \quad (2)$$

The associated uncertainty was obtained by error propagation of the fit parameters:

$$\sigma_{x_{\max}} = \sqrt{\sigma_{x_c}^2 + \left(\frac{\sigma_w}{2}\right)^2} \quad (3)$$

The % of anisotropy ($r$) was calculated by using the following equation applied on background subtracted spectra:

$$r = \frac{|I_{xx} - I_{yy}|}{I_{xx} + I_{yy}} \times 100 \quad (4)$$

where $r$ is % of the anisotropy and $I_{xx}$ and $I_{yy}$ are the intensities of the selected peak along the *a* axis and *b* axis respectively.

**Aknowledgements** N.P. acknowledges the partial funding by the European Union (ERC-CoG, 3DCuT, 101124606), by the Deutsche Forschungsgemeinschaft (DFG 512734967, DFG 492704387, DFG9539383397, DFG 460444718, and DFG 452128813). N.P. aknowledges support from Terra Quantum AG. G.H. acknowledges financial support through start-up funding from IFW Dresden. Work at Brookhaven national laboratory is supported by US DOE (DE-SC0012704). K.N., A.L. and D.W. acknowledge financial support by by the DFG (German Research Foundation) - Project-ID 417590517 - SFB 1415. F.T. and D. M. would like to acknowledge the Project SUPREME - SUPERCONDUCTING EUROPEAN QUANTUM PILOT LINE (HORIZON-JU-CHIPS-2025-QAC1-1-SGA). F.L.S. and G.H. thank Almut Pöhl for the lamellae preparation, Dr. Libo Ma for providing the access to the Raman setup with 458 nm and 660 nm wavelength lasers and Fanzhou Lv for support during the measurements.

**Author contributions** N.P. and G.H. conceived and designed the experiment. F. L. S. performed the experiments and analyzed the data with the contribution of T. C. and S. K. The cuprate crystals were provided by G. G. The Laue measurements were carried out by A. O. and C.G.F.B. The HAADF-STEM measurement was carried out by D.W. The measurements and the results were discussed by F. L. S., T.C., G.H., N. P. and F. L.S., T. C., S. K., D. M., F. T., A.L., K. N., G. H, N. P respectively. F. L. S., N. P., G. H. wrote the manuscript.

**Declaration of Interest** All authors declare no conflict of interest.

[1] K. A. Müller, "The Unique Properties of Superconductivity in Cuprates," Journal of Superconductivity and Novel Magnetism, vol. 27, pp. 2163–2179, 2014.

[2] B. Keimer, S. A. Kivelson, M. R. Norman, S. Uchida, and J. Zaanen, "From quantum matter to high-temperature superconductivity in copper oxides," Nature, vol. 518, pp. 179–186, 2015.

[3] H. Keller, A. Bussmann-Holder, and K. A. Müller, "Jahn–teller physics and high-tc superconductivity," Materials Today, vol. 11, no. 9, pp. 38–46, 2008.

[4] J. H. Pixley and P. A. Volkov, "Twisted nodal superconductors," Annual Review of Condensed Matter Physics, vol. 17, no. Volume 17, 2026, pp. 183–205, 2026.

[5] G.-M. Zhao, "Pairing interactions and pairing mechanism in high-temperature copper oxide superconductors," Phys. Rev. B, vol. 71, p. 104517, 2005.

[6] T. Confalone, F. Lo Sardo, Y. Lee, S. Shokri, G. Serpico, A. Coppo, L. Chirolli, V. M. Vinokur, V. Brosco, U. Vool, D. Montemurro, F. Tafuri, K. Nielsch, G. Haider, and N. Poccia, "Cuprate twistronics for quantum hardware," Advanced Quantum Technologies, vol. 8, no. 11, p. 2500203, 2025.

[7] C. C. Tsuei and J. R. Kirtley, "Pairing symmetry in cuprate superconductors," Rev. Mod. Phys., vol. 72, pp. 969–1016, 2000.

[8] Y. Yu, L. Ma, P. Cai, R. Zhong, C. Ye, J. Shen, G. D. Gu, X. H. Chen, and Y. Zhang, "High-temperature superconductivity in monolayer $Bi_2Sr_{3-x}Ca_xCu_2O_{8+\delta}$," Nature, vol. 575, no. 7781, pp. 156–163, 2019.

[9] H. Ding, T. Yokoya, J. C. Campuzano, T. Takahashi, M. Randeria, M. R. Norman, T. Mochiku, K. Kadowaki, and J. Giapintzakis, "Spectroscopic evidence for a pseudogap in the normal state of underdoped high-tc superconductors," Nature, vol. 352, no. 51-54, 1996.

[10] N. Avraham, B. Khaykovich, Y. Myasoedov, M. Rappaport, H. Shtrikman, D. E. Feldman, T. Tamegai, P. H. Kes, M. Li, M. Konczykowski, K. van der Beek, and E. Zeldov, "'inverse' melting of a vortex lattice," Nature, vol. 411, no. 451-454, 2001.

[11] X. Wang, L. You, D. Liu, C. Lin, X. Xie, and M. Jiang, "Thin-film-like BSCCO single crystals made by mechanical exfoliation," Physica C: Superconductivity, vol. 474,

pp. 13–17, 2012.
[12] P. A. Volkov, S. Y. F. Zhao, N. Poccia, X. Cui, P. Kim, and J. H. Pixley, "Josephson effects in twisted nodal superconductors," Phys. Rev. B, vol. 111, p. 014514, Jan 2025.
[13] T. Confalone, S. Shokri, F. Lo Sardo, V. M. Vinokur, K. Nielsch, G. Haider, and N. Poccia, "Challenges in the electrical engineering of cuprate twistronics," Nature Reviews Electrical Engineering, vol. 2, no. 2, 2025.
[14] S. Y. F. Zhao, X. Cui, P. A. Volkov, H. Yoo, S. Lee, J. A. Gardener, A. J. Akey, R. Engelke, Y. Ronen, R. Zhong, G. Gu, S. Plugge, T. Tummuru, M. Kim, M. Franz, J. H. Pixley, N. Poccia, and P. Kim, "Time-reversal symmetry breaking superconductivity between twisted cuprate superconductors," Science, vol. 382, no. 6677, pp. 1422–1427, 2023.
[15] A. von Hoegen, T. Tai, C. J. Allington, M. Yeung, J. Pettine, M. H. Michael, E. Viñas Boström, X. Cui, K. Torres, A. E. Kossak, B. Lee, G. S. D. Beach, G. D. Gu, A. Rubio, P. Kim, and N. Gedik, "Imaging a terahertz superfluid plasmon in a two-dimensional superconductor," Nature, vol. 650, no. 869-874, 2026.
[16] Y. He, M. Hashimoto, D. Song, S.-D. Chen, J. He, I. M. Vishik, B. Moritz, D.-H. Lee, N. Nagaosa, J. Zaanen, T. P. Devereaux, Y. Yoshida, H. Eisaki, D. H. Lu, and Z.-X. Shen, "Rapid change of superconductivity and electron-phonon coupling through critical doping in Bi-2212," Science, vol. 362, no. 6410, pp. 62–65, 2018.
[17] M. Fratini, N. Poccia, A. Ricci, G. Campi, M. Burghammer, G. Aeppli, and A. Bianconi, "Scale-free structural organization of oxygen interstitials in $La_2CuO_{4+y}$," Nature, vol. 466, p. 841–844, 2010.
[18] F. Massee, Y. K. Huang, and M. Aprili, "Atomic manipulation of the gap in $Bi_2Sr_{3-x}Ca_xCu_2O_{8+\delta}$," Science, vol. 367, no. 6473, pp. 68–71, 2020.
[19] M. A. Subramanian, C. C. Torardi, J. C. Calabrese, J. Gopalakrishnan, K. J. Morrissey, T. R. Askew, R. B. Flippen, U. Chowdhry, and A. W. Sleight, "A new high-temperature superconductor: $Bi_2Sr_{3-x}Ca_xCu_2O_{8+y}$," Science, vol. 239, no. 4843, pp. 1015–1017, 1988.
[20] N. Poccia, G. Campi, M. Fratini, A. Ricci, N. L. Saini, and A. Bianconi, "Spatial inhomogeneity and planar symmetry breaking of the lattice incommensurate supermodulation in the high-temperature superconductor $Bi_2Sr_2CaCu_2O_{8+x}$," Phys. Rev. B, vol. 84, p. 100504(R), 2011.
[21] N. Poccia, S. Y. F. Zhao, H. Yoo, X. Huang, H. Yan, Y. S. Chu, R. Zhong, G. Gu, C. Mazzoli, K. Watanabe, T. Taniguchi, G. Campi, V. M. Vinokur, and P. Kim, "Spatially correlated incommensurate lattice modulations in an atomically thin high-temperature $Bi_2.1Sr_{1.9}CaCu_{2.0}O_{8+x}$ superconductor," Phys. Rev. Mater., vol. 4, p. 114007, Nov 2020.
[22] I. Zeljkovic, E. J. Main, T. L. Williams, M. C. Boyer, K. Chatterjee, W. D. Wise, Y. Yin, M. Zech, A. Pivonka, T. Kondo, T. Takeuchi, H. Ikuta, J. Wen, Z. Xu, G. D. Gu, E. W. Hudson, and J. E. Hoffman, "Scanning tunnelling microscopy imaging of symmetry-breaking structural distortion in the bismuth-based cuprate superconductors," Nature Materials, vol. 11, no. 7, 2012.
[23] J. A. Slezak, J. Lee, M. Wang, K. McElroy, K. Fujita, B. M. Andersen, P. J. Hirschfeld, H. Eisaki, S. Uchida, and J. C. Davis, "Imaging the impact on cuprate superconductivity of varying the interatomic distances within individual crystal unit cells," Proceedings of the National Academy of Sciences, vol. 105, no. 9, pp. 3203–3208, 2008.
[24] Q. Gao, H. Yan, J. Liu, P. Ai, Y. Cai, C. Li, X. Luo, C. Hu, C. Song, J. Huang, H. Rong, Y. Huang, Q. Wang, G. Liu, G. Gu, F. Zhang, F. Yang, S. Zhang, Q. Peng, Z. Xu, L. Zhao, T. Xiang, and X. J. Zhou, "Selective hybridization between the main band and the superstructure band in the $Bi_2Sr_2CaCu_2O_{8+\delta}$ superconductor," Phys. Rev. B, vol. 101, p. 014513, 2020.
[25] T. Honma, K. Yamaya, F. Minami, and S. Takekawa, "Temperature dependences of anisotropic resistivity in the ab plane and hall coefficient in a single-crystal $Bi_2Sr_2CaCu_2O_y$," Physica C: Superconductivity, vol. 176, no. 1, pp. 209–215, 1991.
[26] S. Tajima, G. D. Gu, S. Miyamoto, A. Odagawa, and N. Koshizuka, "Optical evidence for strong anisotropy in the normal and superconducting states in $Bi_2Sr_{3-x}Ca_xCu_2O_{8+z}$," Phys. Rev. B, vol. 48, pp. 16164(R)–16167(R), 1993.
[27] G. Campi, A. Bianconi, N. Poccia, G. Bianconi, L. Barba, G. Arrighetti, D. Innocenti, J. Karpinski, N. D. Zhigadlo, S. M. Kazakov, M. Burghammer, M. v. Zimmermann, M. Sprung, and A. Ricci, "Inhomogeneity of charge-density-wave order and quenched disorder in a high-Tc superconductor," Nature, vol. 525, p. 359–362, 2015.
[28] Y. Toda, F. Kawanokami, T. Kurosawa, M. Oda, I. Madan, T. Mertelj, V. V. Kabanov, and D. Mihailovic, "Rotational symmetry breaking in $bi_2sr_2cacu_2o_{8+\delta}$ probed by polarized femtosecond spectroscopy," Phys. Rev. B, vol. 90, p. 094513, 2014.
[29] B. Gogia, S. C. Kashyap, D. K. Pandya, and K. L. Chopra, "Possible degradation mechanism of BSCCO superconductors," Superconductor Science and Technology, vol. 6, p. 497, jul 1993.
[30] L. J. Sandilands, J. X. Shen, G. M. Chugunov, S. Y. F. Zhao, S. Ono, Y. Ando, and K. S. Burch, "Stability of exfoliated $Bi_2Sr_2CaCu_2O_{8+x}$ studied by raman microscopy," Phys. Rev. B, vol. 82, p. 064503, 2010.
[31] T. Confalone, F. Lo Sardo, D. Montemurro, D. Massarotti, V. M. Vinokur, G. Gu, F. Tafuri, K. Nielsch, G. Haider, and N. Poccia, "Preserving the josephson coupling of twisted cuprate junctions via tailored silicon nitride circuits boards," Small, vol. 21, no. 50, p. e06520, 2025.
[32] M. Martini, Y. Lee, T. Confalone, S. Shokri, C. N. Saggau, D. Wolf, G. Gu, K. Watanabe, T. Taniguchi, D. Montemurro, V. M. Vinokur, K. Nielsch, and N. Poccia, "Twisted cuprate van der waals heterostructures with controlled josephson coupling," Materials Today, vol. 67, pp. 106–112, 2023.
[33] K. Tsutsui, T. Tohyama, and S. Maekawa, "Mott gap excitations and resonant inelastic x-ray scattering in doped cuprates," Phys. Rev. Lett., vol. 91, p. 117001, 2003.
[34] I. Loa, J. Hofmann, A. Litvinchuk, and C. Thomsen, "Superstructure of $Bi_2Sr_2CaCu_2O_8$ superconductors: a raman-scattering study," SPIE, vol. 2696, pp. 306–314, 1996.
[35] M. Boekholt, A. Erle, P. Splittgerber-Hünnekes, and G. Güntherodt, "Phonon raman spectroscopy of superconducting $Bi_2Sr_2CaCu_2O_{8+\delta}$ single crystals," Solid State Communications, vol. 74, no. 10, pp. 1107–1112, 1990.
[36] V. Denisov, B. Mavrin, V. Podobedov, I. Alexandrov,

A. Bykov, A. Goncharov, O. Mel'nikov, and N. Romanova, "Raman spectra of superconducting Bi-Sr-Ca-Cu-O single crystals," Solid State Communications, vol. 70, no. 9, pp. 885–888, 1989.

[37] M. Cardona, C. Thomsen, R. Liu, H. von Schnering, M. Hartweg, Y. Yan, and Z. Zhao, "Raman scattering on superconducting crystals of $Bi_2(Sr_{1-x}Ca_x)_{n+2}Cu_{n+1}O_{(6+2n)+\delta}$ (n = 0, 1)," Solid State Communications, vol. 66, no. 12, pp. 1225–1230, 1988.

[38] R. Liu, M. V. Klein, P. D. Han, and D. A. Payne, "Raman scattering from $A_g$ and $B_{1g}$ phonons in $Bi_2Sr_2Ca_{n-1}Cu_nO_{2n+4}$ (n=1,2)," Phys. Rev. B, vol. 45, pp. 7392–7396, 1992.

[39] A. Martin and M. Lee, "Oxygen isotope effect on the vibrational modes of $Bi_2Sr_2CaCu_2O_{8+\delta}$," Physica C: Superconductivity, vol. 254, no. 3, pp. 222–232, 1995.

[40] I. Figueruelo-Campanero, A. del Campo, G. Nieva, E. M. Gonza´lez, A. Serrano, and M. Menghini, "Apparent color and Raman vibrational modes of the high-temperature superconductor $Bi_2Sr_2CaCu_2O_{8+\delta}$ exfoliated flakes," 2D Materials, vol. 11, no. 2, p. 025032, 2024.

[41] S. Sugai and M. Sato, "Magnon and phonon raman scattering in $Bi_2Sr_2Ca_{1-x}Y_xCu_2O_{8+y}$ ," Phys. Rev. B, vol. 40, pp. 9292–9295, 1989.

[42] M. Boekholt and G. Güntherodt, "Resonant raman scattering of superconducting $Bi_2Sr_2CaCu_2O_{8+\delta}$ single crystals," Physica C: Superconductivity and its Applications, vol. 185, pp. 965–966, 1991.

[43] M. Kakihana, M. Osada, M. Ka¨ll, L. Bo¨rjesson, H. Mazaki, H. Yasuoka, M. Yashima, and M. Yoshimura, "Raman-active phonons in $b_{1g}$ phonons in $Bi_2Sr_2Ca_{1-x}Y_xCu_2O_{8+\delta}$ (x = 0-1): Effects of hole filling and internal pressure induced by y doping for ca, and implications for phonon assignments," Phys. Rev. B, vol. 53, pp. 11796–11806, 1996.

[44] S. Shokri, M. Ceccardi, T. Confalone, C. N. Saggau, Y. Lee, M. Martini, G. Gu, V. M. Vinokur, I. Pallecchi, K. Nielsch, F. Caglieris, and N. Poccia, "Evolution of dissipative regimes in atomically thin $Bi_2Sr_2CaCu_2O_{8+x}$ superconductor," Advanced Electronic Materials, vol. 11, no. 4, p. 2400496, 2025.

[45] Y. Lee, M. Martini, T. Confalone, S. Shokri, C. N. Saggau, D. Wolf, G. Gu, K. Watanabe, T. Taniguchi, D. Montemurro, V. M. Vinokur, K. Nielsch, and N. Poccia, "Encapsulating high-temperature superconducting twisted van der waals heterostructures blocks detrimental effects of disorder," Adv. Materials, vol. 35, 2023.

[46] S. Abdelhaleem, M. O. Alziyadi, A. Alruwaili, M. J. Alawi, A. Alkabsh, and M. S. Shalaby, "BSCCO high $T_c$ - superconductor materials: strategies toward critical current density enhancement and future opportunities," Applied Physics A, vol. 131, no. 2, 2025.

[47] O. Can, T. Tummuru, R. P. Day, I. Elfimov, A. Damascelli, and M. Franz, "High-temperature topological superconductivity in twisted double-layer copper oxides," Nature Physics, vol. 17, pp. 519–524, 2021.

[48] H. Jin, G. Serpico, Y. Lee, T. Confalone, C. N. Saggau, F. Lo Sardo, G. Gu, B. H. Goodge, E. Lesne, D. Montemurro, K. Nielsch, N. Poccia, and U. Vool, "Exploring van der Waals Cuprate Superconductors Using a Hybrid Microwave Circuit," Nano Letters, vol. 25, pp. 3191–3198, 01 2025.

[49] C. N. Saggau, S. Shokri, M. Martini, T. Confalone, Y. Lee, D. Wolf, G. Gu, V. Brosco, D. Montemurro, V. M. Vinokur, K. Nielsch, and N. Poccia, "2D High - Temperature Superconductor Integration in Contact Printed Circuit Boards," ACS Applied Materials & Interfaces, vol. 15, no. 44, pp. 51558–51564, 2023.

[50] K. Moffat, "Laue diffraction," in Macromolecular Crystallography Part B, vol. 277 of Methods in Enzymology, pp. 433–447, Academic Press, 1997.

[51] F. Hofmann, X. Song, T.-S. Jun, B. Abbey, M. Peel, J. Daniels, V. Honkima¨ki, and A. Korsunsky, "High energy transmission micro-beam laue synchrotron x-ray diffraction," Materials Letters, vol. 64, no. 11, pp. 1302–1305, 2010.

[52] F. Lo Sardo, M. Esposito, T. Confalone, C. Tremblay, V. M. Vinokur, G. Gu, D. Montemurro, D. Massarotti, F. Tafuri, K. Nielsch, N. Poccia, and G. Haider, "Quantifying Twist Angles in Cuprate Heterostructures with Anisotropic Raman Signatures," Advanced Physics Research, vol. 4, no. 12, p. e00108, 2025.

[53] C. Yelpo, R. Faccio, D. Ariosa, and S. Favre, "Electronic and vibrational properties of the high tc superconductor $Bi_2Sr_2CaCu_2O_8$: an ab initio study," Journal of Physics: Condensed Matter, vol. 33, no. 18, p. 185705, 2021.

[54] T. Konstantinova, J. D. Rameau, A. H. Reid, O. Abdurazakov, L. Wu, R. Li, X. Shen, G. Gu, Y. Huang, L. Rettig, et al., "Nonequilibrium electron and lattice dynamics of strongly correlated bi2sr2cacu2o8+ $\delta$ single crystals," Science advances, vol. 4, no. 4, p. eaap7427, 2018.

[55] T. P. Devereaux and R. Hackl, "Inelastic light scattering from correlated electrons," Rev. Mod. Phys., vol. 79, pp. 175–233, 2007.

[56] E. Faulques, "Raman Spectra of Unconventional Superconductors," Materials Synthesis and Characterization, pp. 61–102, 1997.

[57] E. Y. Sherman and C. Ambrosch-Draxl, "Resonant Raman scattering in the superconducting cuprates: Frozen-phonon versus perturbational approach," The European Physical Journal B - Condensed Matter and Complex Systems, vol. 16, no. 2, pp. 251–259, 2000.